\documentclass[a4paper]{PoS}
\usepackage{amssymb,amsmath, fontenc, times, mathptmx, graphicx} 

\title{Composite Higgs Models: a new holographic approach}

\ShortTitle{Composite Higgs Models: a new holographic approach}

\author{\speaker{Alisa Katanaeva} and Dom\`enec Espriu \footnote{We acknowledge financial support from the following grants:  FPA2013-46570-C2-1-P (MINECO), 2014SGR104 
(Generalitat de Catalunya), and MDM-2014-0369.}\\
        Departament de F\'\i sica Qu\`antica i Astrof\'\i sica and \\
Institut de Ci\`encies del Cosmos (ICCUB), Universitat de Barcelona,\\ 
Mart\'i i Franqu\`es 1, 08028 Barcelona, Catalonia, Spain\\
        E-mail: \email{katanaeva@fqa.ub.edu}, \email{espriu@icc.ub.edu}}

\abstract{We revisit the construction of the composite Higgs models in a context of the bottom-up holographic approach. The soft wall framework is under consideration imposing the translation of the $4D$ global symmetry breaking characteristic to the new strongly interacting sector in the $5D$ bulk.  The focus stays on the minimal $SO(5)\to SO(4)$ breaking pattern. 
The $5D$ model has a specific form inspired by the effective models of QCD, representing a generalized sigma model coupled both to the composite resonances and to the SM gauge bosons. The latter are treated as external $4D$ sources and conceptually develop no propagation into the bulk.
The holographic description allows for the consideration of spin one and spin zero resonances.
The resulting spectrum leads in a natural way to a variety of new composite resonances, four of which represent 
the massless Goldstone bosons. Existing experimental constraints
on the electroweak precision parameters permit to accommodate vector and scalar resonances with masses in the $1 - 2$ TeV range without difficulties, but higher masses are possible too.
Moreover, for the SM gauge fields holography provides relevant vacuum polarization amplitudes and mixing with composite resonances. 
Further considering higher order correlation functions we can formulate semi-quantitative predictions for the effective couplings and cross-sections. These provide additional restrictions that are currently being investigated.
}

\FullConference{XIII Quark Confinement and the Hadron Spectrum - Confinement2018\\
		31 July - 6 August 2018\\
		Maynooth University, Ireland}

\newcommand{\be}{\begin{equation}}
\newcommand{\ee}{\end{equation}}
\newcommand{\ba}{\begin{array}}
\newcommand{\eaq}{\end{array}}
\newcommand{\bea}{\begin{eqnarray}}
\newcommand{\eea}{\end{eqnarray}}

\newcommand{\nn}{\nonumber}

\newcommand{\bi}{\begin{itemize}}
\newcommand{\ei}{\end{itemize}}
\newcommand{\bal}{\begin{aligned}}
\newcommand{\eal}{\end{aligned}}
\newcommand{\Tr}{\operatorname{Tr}}
\newcommand{\Exp}{\operatorname{Exp}}

\begin{document}

\section{Introduction}
The idea of compositeness of the Higgs boson is motivated by the fact that the minimal version of the Standard Model (SM) with one Higgs doublet although successfully incorporating the electroweak symmetry breaking (EWSB) does not provide a full understanding of the scales involved nor any explanation on the origin of the fundamental scalar itself. 

In the composite models of Georgi and Kaplan one considers an additional strongly interacting sector, 
characterized by the number of (techni)colours $N_{tc}$,
with a global group $G$ describing a (techni)flavour symmetry. The group $G$ is broken at a scale $\Lambda_{UV}$
to its subgroup $H$ due to some unspecified QCD-like forces; $SU(2)\times U(1)$ group must remain in the unbroken sector.
Composite Goldstone bosons appear in the coset $G/H$, and the states with the quantum numbers of the Higgs boson should be present 
among them. 
The SM group itself $SU(2)'\times U(1)'$ lies in $H'$, which is rotated with respect to $H$ 
around one of the broken directions by a misalignment angle $\theta$.
Thus, the degree of the breaking of the weak interactions becomes an alignment issue
and $\theta$ establishes an hierarchy between the UV, $\Lambda_{UV}=4\pi F$, and Fermi scales, $\Lambda_{IR}=4\pi v$,
via the relation $F=v\sin\theta$.

Further we focus on the minimal composite Higgs model (MCHM) with $SO(5)\to SO(4)$ breaking pattern \cite{ACP_2005}. 
That is the  minimal structure to preserve the custodial symmetry and have exactly one Higgs doublet in the coset.
In the same time one cannot reproduce an $SO(5)$ global symmetry at the microscopic level introducing fundamental 
(techni)fermions \cite{Cacciapaglia:2014}. The experimental bound on the misalignment in the conventional MCHM is $\sin\theta \leq 0.34$~\cite{Atlas_2015}, assuming the coupling  of the Higgs to gauge bosons $\kappa_v = \cos\theta$, even though this identification will need to be revised in the context of this type of models.

The holographic technique provides a way to analyze the impact of the compositeness hypothesis on some observables 
due to facilitation of the calculations related to the strongly interacting sector. 
Strictly speaking the obtained results correspond to the large $N_c$ (or $N_{tc}$) limit, but it is common
to make a phenomenological sidestep towards some finite values. Moreover it is not possible to proceed
 model-independently and some specifications are necessary. We propose to use the bottom-up holographic approach
with a soft wall (SW). In \cite{ACP_2005} and subsequent works the hard wall (HW) option was generally used, with
the breaking realized on the IR brane. The SW in composite Higgs context is much less studied (there is only \cite{Falkowski2008})
for no apparent reason as, for instance, in QCD it proved to give better description of the meson phenomenology \cite{SW_2006}.
The breaking in SW scenario also appears in the $5D$ bulk implying a more complicated structure of the $5D$ Lagrangian, 
including vector and scalar states associated to broken and unbroken sectors.

Another issue is the introduction of the SM gauge bosons. That is usually performed in the bulk imposing a Neumann condition
in the UV for some modes (as in \cite{Falkowski2008}). However, in our opinion, this promotion of EW bosons into the bulk
comes in tension with the holographic treatment being supposedly valid only in the regime of a strong coupling.
We propose to treat SM gauge fields perturbatively on the UV brane and consider them as sources of the
vector currents of $SO(5)$ with the same quantum numbers, thus
\be\label{lagr1}
\mathcal L_{4D} = \mathcal {\widetilde L}_{str. int.} + \mathcal L_{SM} + \widetilde J^{a\ \mu}  W_\mu^{a}
+ \widetilde J^{Y\ \mu}  B_\mu,
\ee
where the tilde on the Lagrangian of the strongly interacting sector and its currents ($J^{a}_\mu$ and $J^{Y}_\mu$)
signifies the realization of the misalignment through the rotation of the $SO(5)$ generators,
\be
T^A(\theta)=r(\theta)T^A(0)r^{-1}(\theta), \ \text{with} \ r(\theta)=\begin{pmatrix}
          1_{3\times3} & 0 & 0\\
          0 &\cos(\theta)&\sin(\theta)\\
          0 & -\sin(\theta)&\cos(\theta)\\
         \end{pmatrix}, \quad A=1,...,10.
\ee

In order to have less free parameters it is essential to make an assumption on the microscopic
structure of the strongly interacting sector. This can be achieved by constructing the two-point correlators of the following operators and matching their short-distance expansion to a holographic result.
We choose a scalar field of rank 2, $s^{\alpha\beta}$, then the Lagrangian invariant under the global 
$SO(5)$ transformation
is $\mathcal L =\frac12 \partial_\mu s_{\alpha\beta}\partial^\mu s^\top_{\beta\alpha}
-\frac12 m^2 s_{\alpha\beta}s^\top_{\beta\alpha}$. We can construct a scalar invariant 
$s^{\alpha\gamma}s^{\gamma\alpha}$, giving a scalar operator 
$\mathcal O_{sc}^{\alpha\beta}(x)=s^{\alpha\gamma}s^{\gamma\beta}$ with dimension $\Delta=2$, spin $p=0$; and a Noether current 
$i[T^A,s]_{\alpha\beta}\partial^\mu s^\top_{\beta\alpha}$ giving a vector operator $\mathcal O_{vec}^{A\ \mu}(x)$, with $\Delta=3$, $p=1$.
We establish the $5D$ masses for the fields dual to these operators following the general formula from the AdS/CFT dictionary 
$M^2R^2=(\Delta-p)(\Delta+p-4)$ \cite{Maldacena_1999,Gubser1998,Witten_1998}.

In the end holography provides all the necessary $n$-point functions of the composite operators to 
calculate self-energies for the SM gauge bosons and analyze possible effective interactions and mixings
between EW and composite degrees of freedom.

\section{$5D$ setup}
We work with the $5D$ AdS metric given by $g_{MN}dx^M dx^N=\frac{R^2}{z^2}(\eta_{\mu\nu}dx^\mu dx^\nu-d^2z),$
where $R$ is the AdS radius and the convention for the Minkowski space is $\eta_{\mu\nu}=\text{diag}(1,-1,-1,-1)$.
The $SO(5)$ invariant action has the following form 
\begin{align}\label{5Daction}
S_{5D}=&-\frac{1}{4g_5^2}\int d^4xdz\sqrt{-g}e^{-\Phi(z)}\Tr F_{MN}F_{MN} \\ \nn
&+\frac1{k_s}\int d^4xdz\sqrt{-g}e^{-\Phi(z)}\Tr\bigg[D_M H^\top D^M H-M^2 HH^\top- M^2 (HD^\top+H^\top D) \bigg]. \label{5Daction}
\end{align}
The normalization constants have the dimensionality $[g_5^2]=[k_s]=L^1$; and following the SW holographic approach we have introduced a dilaton exponent with $\Phi(z)$.
The $5D$ mass of the scalar field $H(x,z)$ is $M^2R^2=-4$, while the vector fields $A_M$ get the zero one.
The dynamical breaking from $SO(5)$ to $SO(4)$ happens in the scalar sector due to a function $f(z)$ appearing in the nonlinear parameterization of $H$:
\be
H(x,z)=\xi\Sigma\xi^{-1},\quad \Sigma(x,z)=\begin{pmatrix}
          0_{4\times4} & 0\\
          0 & f(z)\\
         \end{pmatrix}+iT^a\sigma^a(x,z),
         \quad \xi(x,z)=\exp\left(\frac{i\pi^i(x,z)\widehat{T}^i}{\sqrt 2 f(z)}\right).
\ee
We use a standard representation of the $SO(5)$ generators enumerating separately the ones of the unbroken $SO(4)$ sector as $T^a, \ a=1,...,6$ and
the rest which are broken $\widehat{T}^i,\ i=1,...,4$. Consequently, for the vector fields we have $A_M=A_M^AT^A=A_M^aT^a+A_M^i\widehat{T}^i$.

The matrix field $D$ is introduced in Eq.~(\ref{5Daction}) to provide an explicit soft breaking that is used in order to fine-tune to zero the masses of the would-be Goldstone bosons $\pi^i$, as the boundary conditions make them naturally massive. It is parameterized by a function $b(z)$ as $ D=\begin{pmatrix}
		0_{4\times4} & 0\\
		0 &  b (z)\\
		\end{pmatrix}$.
		
The summary of ans\"atze functions is: $\Phi(z)=\kappa^2z^2$, $f(z)=f\cdot\kappa z$, $b(z)/f(z)=\mu_1+\mu_2\cdot\kappa z,$ where
we determine $\mu_1=\mu_2=-1$ (massless condition for the Goldstone bosons), while $f$ and $\kappa$ are the parameters of the model.
We choose the $A_z=0$ gauge which is standard for SW models in QCD and can still set $\partial^\mu A_\mu =0$ with a consistent gauge transformation.

From the quadratic part of the $5D$ Lagrangian we can get the masses of the composite resonances in $4D$ and the two-point
correlators of the composite operators. For the properties of the $4D$ resonances we look for the normalizable solutions
of equations of motion subject to the Dirichlet boundary condition at $z=\varepsilon$. For the correlators we can use the 
AdS/CFT prescription on the variation of the on-shell $5D$ action that is holographically connected to the $4D$ partition
function $\ln \mathcal Z_{4D}$ defined via
\be
\mathcal Z_{4D}[\phi_{\mathcal O}]=\int [\mathcal D s]\Exp i\int d^4x [\mathcal L_{str. int.}(x)+ 
\phi_{\mathcal O \mu}^A(x)\mathcal O_{vec}^{A\ \mu}(x)
+\phi_{\mathcal O}^{\alpha\beta}(x)\mathcal O_{sc}^{\beta\alpha}(x)].\label{Zqft}
\ee

Consider first the vector sector. The $5D$ vector fields can be represented as Kaluza--Klein (KK)  infinite towers 
of $4D$ massive states with specific $z$-profiles holographically provided. The $4D$ masses explicitly depend on the model,
and in the present context we have ($V$ and $A$ correspond to unbroken and broken directions in resemblance 
to vectors and axial-vectors of QCD):
\be
M_V^2(n)=4\kappa^2(n+1),\quad M_A^2(n)=4\kappa^2\left(n+1+\frac{(g_5Rf)^2}{2k_s}\right),\quad n=0,1,2,....
\ee

The vector correlators are defined as follows 
	\begin{eqnarray}
 	&\langle \mathcal O_\mu^{a/i}(q) \mathcal O_\nu^{b/j}(p)\rangle=\delta(p+q)\int d^4x e^{iqx}\langle \mathcal O_\mu^{a/i}(x) \mathcal O_\nu^{b/j}(0)\rangle
	=\frac{\delta^2 iS_{5D}^{on-shell}}{\delta i \phi^{a/i}_{\mathcal O\mu}(q)\delta i \phi^{b/j}_{\mathcal O\nu}(p)},\\ \notag
	&i\int d^4x e^{iqx}\langle \mathcal O_\mu^{a/i}(x) \mathcal O_\nu^{b/j}(0)\rangle=\delta^{ab/ij}\left(\frac{q_\mu q_\nu}{q^2}-\eta_{\mu\nu}\right)\Pi_{V/A}(q^2).
	\end{eqnarray}
After subtracting the generic ambiguities of a form $C_0+C_1q^2$ we obtain the following result
\begin{align}
 \Pi_{V}(q^2)=\sum\limits_{n}\frac{q^4F_V^2}{M^2_V(n)(-q^2+M^2_V(n))},\quad 
 \Pi_{A}(q^2)=\sum\limits_n\frac{q^4 F_A^2(n)}{M^2_A(n)(-q^2+M^2_A(n))}-F^2;\\
 F_V^2=\frac{2R\kappa^2}{g_5^2},\quad F_A^2(n)=\frac{2R\kappa^2}{g_5^2}\frac{n+1}{n+1+\frac{(g_5Rf)^2}{2k_s}},\quad 
F^2=\frac{2R\kappa^2}{g_5^2}\sum\limits_n\frac{\frac{(g_5Rf)^2}{2k_s}}{n+1+\frac{(g_5Rf)^2}{2k_s}}
\end{align}

A similar analysis applies for the part of the Lagrangian with the scalar fields. The masses of the KK radial
excitations in the unbroken scalar and broken Goldstone sectors are
\be
 M_\sigma^2(n)=4\kappa^2(n+1),\quad M_\pi^2(n)=4\kappa^2 n,\quad n=0,1,2,....
\ee
And for the correlators we can get (the proper operators descend from $\mathcal O_{sc}^{\alpha\beta}$: $\mathcal O_s^a = T^ass,\ 
\mathcal O_p^a=\widehat T^ass$)
\begin{align}
&i\int d^4x e^{iqx}\langle \mathcal O_{s/p}^a(x) \mathcal O_{s/p}^b(0)\rangle=\delta^{ab}\Pi_{S/G}(q^2),\\
\Pi_S(q^2)=\sum\limits_{n}\frac{F^2_\sigma}{q^2-M^2_\sigma(n)},&\
\Pi_G(q^2)=\sum\limits_{n}\frac{F^2_\pi}{q^2-M^2_\Pi(n)};\quad F^2_\sigma=\frac{16\kappa^2R}{k_s} ,\
F^2_\pi=\frac{16\kappa^2R}{k_s}.
\end{align}

The free parameters $g_5^2$ and $k_s$ can be matched to a single parameter of the $4D$ strongly interacting sector. 
The large $Q^2$ limit of the listed correlators should be compared with the one obtained by  the usual field theory methods 
in $4D$. We find that the following relations are valid:
\be
\frac{k_s}{R}=\frac{64 \pi^2}{5N_{tc}},\quad \frac{g_5^2}{R}=\frac{8 \pi^2}{5N_{tc}}.
\ee

\section{Two point functions and mixings for the SM bosons}

In the effective Lagrangian~(\ref{lagr1}) a certain $SU(2)'\times U(1)'\subset SO(4)'$ is already gauged
because the SM fields $W_\mu^{a}$ and $B_\mu$  couple to the particular currents of the strongly interacting sector. 
 They are among the vector currents that are holographically connected to the vector composite fields. Let us name
 the first three operators of the unbroken vector sector $\mathcal O_\mu^{a_L}(x)$ and the last three -- $\mathcal O_\mu^{a_R}(x)$.
 Then $W_\mu^a$ couples to $J^a_\mu=\frac g{\sqrt2} O_\mu^{a_L}$ and $B_\mu$ to $J^Y_\mu=\frac {g'}{\sqrt2} O_\mu^{3_R}$, 
 as we assume the hypercharge to be realised as $Y=T_{3_R}$.
Hence, we may include to the $4D$ partition function the following terms quadratic in the external sources $W$ and $B$:
${W}^\mu\langle \widetilde J^L_\mu(q) \widetilde J^L_\nu(-q)\rangle {W}^\nu $,
${W}^\mu  \langle \widetilde J^L_\mu(q) \widetilde J^R_\nu(-q)\rangle {B}^\nu $,
${B}^\mu  \langle \widetilde J^R_\mu(q) \widetilde J^R_\nu(-q)\rangle {B}^\nu $.
 The relevant quadratic contribution of the $4D$ effective action is
\begin{align}
\notag
S_{4D}^{eff}\supset\int d^4q\bigg[&\left(\frac{q^\mu q^\nu}{q^2}-
\eta^{\mu\nu}\right)\frac{1}{4}\Pi_{diag}(q^2)(g^2{W}_\mu^{1}{W}_\nu^{1}+g^2{W}_\mu^2{W}_\nu^2+g^2{W}_\mu^3{W}_\nu^3+g'^2{B}_\mu{B}_\nu)
\\ \label{gaugedL}
&+\left(\frac{q^\mu q^\nu}{q^2}-\eta^{\mu\nu}\right)\frac{1}4 \Pi_{LR}(q^2) gg'{W}_\mu^{3}{B}_\nu\bigg],\\
 &\Pi_{diag}(q^2)=\frac{1+\cos^2\theta}2 \Pi_{V}(q^2)+\frac{\sin^2\theta}2 \Pi_{A}(q^2),\\
&\Pi_{LR}(q^2)=\sin^2\theta \left(\Pi_{V}(q^2)-\Pi_{A}(q^2)\right)
\end{align}
The diagonal self-energies result in the mass terms for the gauge fields in a small $q^2$ limit
\be
M^2_W=\frac{g^2}4\sin^2\theta F^2,\quad M^2_Z=\frac{g^2+g'^2}4\sin^2\theta F^2,\quad M^2_\gamma=0. \label{Gmasses}
\ee
The left-right two-point function defines the $S$ parameter of Peskin and Takeuchi.
In terms of the masses and decay constants
of the vector composite states it gets a form  (in our description $F_V(n)=F_V$ for all values of $n$)
\be \label{Spar}
S=4\pi\sin^2\theta \bigg[\sum\limits_n\frac{F^2_V(n)}{M^2_V(n)}-\sum\limits_n\frac{F^2_A(n)}{M^2_A(n)}\bigg].
\ee
All other electroweak oblique parameters are vanishing or naturally small in the considered model.

At the same time the structure of the correlation functions provides the mixing between gauge bosons and composite resonances.
For instance, for the $W$ field we have ($\mathcal D^{\mu\nu}=\Box\eta^{\mu\nu}-\partial^\mu\partial^\nu$)
\begin{align}\nn
+\frac g{\sqrt2}W_\mu^a(x) \mathcal D^{\mu\nu}&\sum\limits_n \frac{F_V}{M_V(n)}  \left[\frac{1+\cos\theta}2 A_{L\ \nu(n)}^a(x) +
\frac{1-\cos\theta}2 A_{R\ \nu(n)}^a(x)\right]\\ 
-\frac g{\sqrt2} W_\mu^a(x)\mathcal D^{\mu\nu}&\sum\limits_n \frac{F_A(n)}{M_A(n)} \frac{\sin\theta}{\sqrt2} A_{br\ \nu(n)}^a(x)  
\end{align}

It is straightforward to get the eigenstates subsequently diagonalizing the kinetic terms and mass matrices. We show further that the mixing is not very significant numerically. 
\section{Mass and mixing estimations}
\begin{figure}[t]
	\centerline{\includegraphics[scale=0.5]{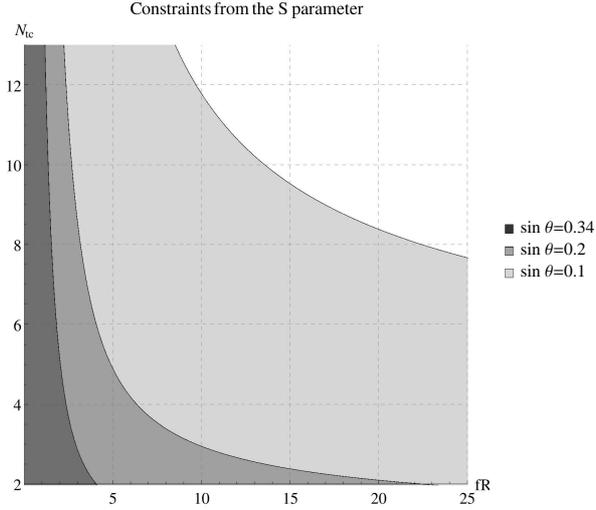}}
	\caption{\label{Spar_fig} The $(\sin\theta, fR, N_{tc})$ parameter region allowed by the $S$ parameter restraints.}
\end{figure}

\begin{table}[b]\center
\caption{Different predictions of the minimal vector masses for $\sin\theta=0.25$ and $0.30$.}
{\begin{tabular}{@{}ccc|c|c@{}} 
$\sin\theta$ & $N_{tc}$ & $fR$  & $M_V(0)$, $\text{TeV}$ & $M_A(0)$, $\text{TeV}$ \\
\hline
$0.25$& $2$ & $9.1$ & $0.89$ & $2.20$  \\
$0.25$& $3$ & $5.2$ & $1.21$ & $1.99$  \\
$0.25$& $4$ & $3.9$ & $1.37$ & $1.92$ \\
$0.25$& $10$ & $2.0$ & $1.66$ & $1.86$  \\
\hline
$0.30$& $2$ & $5.5$ & $1.26$ & $2.14$  \\
$0.30$& $3$ & $3.7$ & $1.50$ & $2.03$ \\
$0.30$& $4$ & $2.9$ & $1.61$ & $1.99$ \\
$0.30$& $10$ & $1.6$ & $1.81$ & $1.96$ \\
\end{tabular}\label{tab-0}}
\end{table}
The masses of the composite resonances are governed by the scale of parameter $\kappa$. The latter can be constrained from the experimental values known for quantities of Eqs.~(\ref{Gmasses}) and (\ref{Spar}). The first provides a particular equation that combines the model parameters $\kappa,\ \sin\theta,\ fR,\ N_{tc}$.  The latter should be considered as an expression of the $S$ parameter in terms of $\sin\theta,\ fR,\ N_{tc}$. We take from \cite{Gfitter_2014} 
\be
-0.06\leq S\leq0.16,\label{ph1}
\ee
and obtain the areas these parameters may span as depicted in Fig.~\ref{Spar_fig}.
Further, we would like to see the extreme case, {\it i.e.} to take the parameters maximally saturating the $S$ bound. This results in the minimal possible values of the masses. One can see from Table~\ref{tab-0} that the proposed programme can accommodate rather light values for the ground states $M_V(0)$ and $M_A(0)$ of order $1-2$~TeV, but higher masses are certainly not excluded.

Other general observations are following. Consider fixing any two parameters among $(\sin\theta,$ $fR,\ N_{tc})$, then the growth of the third parameter results in smaller $\kappa$ and a possibility of lower masses. Indeed, an unlimited growth in $fR$ results in unlikely small masses for $\sin\theta \lesssim 0.1$. However, higher values of other two parameters soon face the upper experimental limit of the $S$ parameter.


Concerning the mixing we take some particular cases corresponding to the lines in Table~\ref{tab-0}. In our estimation the main demonstration of the mixing is in a state corresponding to the diagonal '$W$'; it may have from $2$ to $6\ \%$ of $A_L$ and the rest of original $W$. The changes are rather insignificant: only the diagonal '$A_L$' gets $1-2\ \%$ heavier. However, these results are for the minimal mass option argued above. One can consider heavier states taking lower values of $fR$ for the same $\sin\theta$ and $N_{tc}$, then the mixing is more obvious, {\it e.g.} between $A_L$ and $A_{br}$. The masses of diagonal '$A_L$' and '$A_{br}$' tend to become larger, while those of '$A_R$' and '$W$' remain the same.
\section{Three point functions}
\begin{figure}[t]\center
 \includegraphics[width=0.5\textwidth]{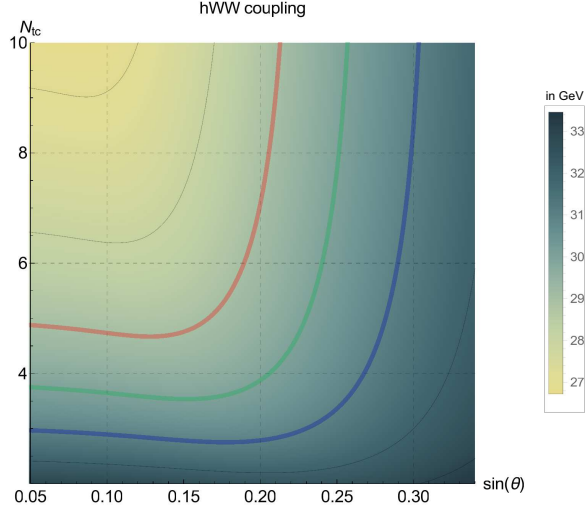}
 \caption{Values of the $hWW$ coupling on the $N_{tc} - \sin \theta$ plane. The curves show the particular constant values with a step of $1$~GeV, in particular, the red  one represents $29$ GeV, green - $30$ GeV and blue - $31$ GeV.  \label{hww_coupl}}
\end{figure}

We focus now in some particular  couplings. Let us consider {\it e.g.} the$g_{hWW}$ vertex. To get the direct coupling of this kind we modify the $5D$ 
Lagrangian with a redefined covariant derivative
\be
D_\mu H(x,z)=\partial_\mu H(x,z)-i[A_\mu(x,z), H(x,z)]-i[\widetilde{X}_\mu(x), H(x,z)]
\ee
where we include gauge boson fields in a bulk through a rotated field with $SO(4)$ indices $\widetilde{X}_\mu=X_\mu^{a}r^{-1}(\theta)T^{a}(0)r(\theta)$. It is important that $X_\mu$ is assumed to be $z$-independent. 
The particular connection to the SM $W$ is $X_\mu^{L\ \alpha}=\frac g{\sqrt2} W_\mu^{\alpha}$.
This modification results in a particular three point vertex $WW\pi^4$ in $5D$. The $4D$ vertex is obtained integrating out the $z$-dimension using the Kaluza--Klein representation of $\pi^4$ and flat profiles of $W$. Recognizing $n=0$ mode of $\pi^4$ as the Higgs field we get
\be
\mathcal L_{4D}\supset\frac{g_{hWW}}2 h W^{1,2}_{\mu}W^{1,2\mu},\quad 
g_{hWW}=\frac{g^2\sin\theta}2 \kappa (fR) \sqrt{\frac R{k_s}} \cdot \sqrt{\frac\pi2}\cos\theta.
\ee
The possible value of the coupling are depicted at Fig.~\ref{hww_coupl}, $\kappa$ parameter is determined following the minimal mass approach of the previous section.
These results give couplings well below the naive estimation $\kappa_V=cos\theta$. Rather, the comparison with the SM analytic expression $g^{SM}_{hWW}=\frac{g^2F\sin\theta}2$ shows that in the present model
\be
g_{hWW}=g^{SM}_{hWW}\cdot \kappa_V, \quad 
\kappa_V=\cos\theta\sqrt{\frac\pi2}\left(\sum_n\frac1{1+n+\frac{(g_5Rf)^2}{2k_s}}\right)^{-1/2}.
\ee

Additionally there is a contribution to $g_{hWW}$ from the mixing of $W$ with vector resonances. As well through the mixing there exist $WW$-composite resonance effective vertices that could be considered. A more detailed analysis will be given in a future publication.

\section{Conclusions}

A new $5D$ holographic setup for the description of the composite Higgs phenomena is presented here. It is inspired by AdS/QCD achievements but has a distinct Lagrangian of a generalized sigma model coupled both to the composite resonances and to the SM gauge bosons.
The particular ansatz consists in the dilaton $z$-profile (common to all SW holographic models), and two functions $f (z)$ and $b(z)$.
Among the model features we highlight that the Goldstone bosons can be made exactly massless;
the vectors and scalars of the unbroken sector are degenerate in mass; $g_{hWW}$ is distinct from that of the conventional MCHM.

From the phenomenological point it is significant that the $S$ parameter restrictions can be met in quite large areas in the parameter space, where in the same time a resonance between $1$ and $2$ TeV could be accommodated.

\bibliography{biblio/composite_higgs_database}

\providecommand{\href}[2]{#2}\begingroup\raggedright\begin{thebibliography}{1}

\bibitem{ACP_2005}
K.~Agashe, R.~Contino and A.~Pomarol, \emph{{The Minimal Composite Higgs
  Model}}, \href{http://dx.doi.org/10.1016/j.nuclphysb.2005.04.035}{\emph{Nucl.
  Phys. B} {\bfseries 719} (2005) 165--187},
  [\href{https://arxiv.org/abs/hep-ph/0412089}{{\ttfamily hep-ph/0412089}}].

\bibitem{Cacciapaglia:2014}
G.~Cacciapaglia and F.~Sannino, \emph{{Fundamental Composite (Goldstone) Higgs
  Dynamics}}, \href{http://dx.doi.org/10.1007/JHEP04(2014)111}{\emph{J. High
  Energy Phys.} {\bfseries 04} (2014) 111},
  [\href{https://arxiv.org/abs/1402.0233}{{\ttfamily 1402.0233}}].

\bibitem{Atlas_2015}
{\scshape The ATLAS collaboration} collaboration, G.~Aad et~al.,
  \emph{Constraints on new phenomena via {H}iggs boson couplings and invisible
  decays with the {ATLAS} detector},
  \href{http://dx.doi.org/10.1007/JHEP11(2015)206}{\emph{J. High Energy Phys.}
  {\bfseries 2015} (nov, 2015) 206},
  [\href{https://arxiv.org/abs/1509.00672v2}{{\ttfamily 1509.00672v2}}].

\bibitem{Falkowski2008}
A.~Falkowski and M.~Perez-Victoria, \emph{{Electroweak Breaking on a Soft
  Wall}}, \href{http://dx.doi.org/10.1088/1126-6708/2008/12/107}{\emph{J. High
  Energy Phys.} {\bfseries 2008} (2008) 107},
  [\href{https://arxiv.org/abs/0806.1737}{{\ttfamily 0806.1737}}].

\bibitem{SW_2006}
A.~Karch, E.~Katz, D.~T. Son and M.~A. Stephanov, \emph{{Linear Confinement and
  AdS/QCD}}, \href{http://dx.doi.org/10.1103/PhysRevD.74.015005}{\emph{Phys.
  Rev. D} {\bfseries 74} (2006) 015005},
  [\href{https://arxiv.org/abs/hep-ph/0602229v2}{{\ttfamily
  hep-ph/0602229v2}}].

\bibitem{Maldacena_1999}
J.~Maldacena, \emph{{The Large N Limit of Superconformal Field Theories and
  Supergravity}}, \href{http://dx.doi.org/10.1023/a:1026654312961}{\emph{Int.
  J. Theor. Phys.} {\bfseries 38} (1999) 1113--1133},
  [\href{https://arxiv.org/abs/hep-th/9711200}{{\ttfamily hep-th/9711200}}].

\bibitem{Gubser1998}
S.~Gubser, I.~Klebanov and A.~Polyakov, \emph{Gauge theory correlators from
  non-critical string theory},
  \href{http://dx.doi.org/10.1016/S0370-2693(98)00377-3}{\emph{Phys. Lett. B}
  {\bfseries 428} (1998) 105 -- 114},
  [\href{https://arxiv.org/abs/hep-th/9802109}{{\ttfamily hep-th/9802109}}].

\bibitem{Witten_1998}
E.~Witten, \emph{{Anti De Sitter Space And Holography}},
  \href{http://dx.doi.org/10.4310/ATMP.1998.v2.n2.a2}{\emph{Adv. Theor. Math.
  Phys.} {\bfseries 2} (1998) 253--291},
  [\href{https://arxiv.org/abs/hep-th/9802150}{{\ttfamily hep-th/9802150}}].

\bibitem{Gfitter_2014}
{\scshape The Gfitter Group} collaboration, M.~Baak, J.~Cuth, J.~Haller,
  A.~Hoecker, R.~Kogler, K.~Moenig et~al., \emph{The global electroweak fit at
  {NNLO} and prospects for the {LHC} and {ILC}},
  \href{http://dx.doi.org/10.1140/epjc/s10052-014-3046-5}{\emph{Eur. Phys. J.
  C} {\bfseries 74} (sep, 2014) 3046},
  [\href{https://arxiv.org/abs/1407.3792v1}{{\ttfamily 1407.3792v1}}].

\end{thebibliography}\endgroup

\end{document}